\pdfoutput=1 
\documentclass[conference]{IEEEtran}
\usepackage{amsmath}
\usepackage{url}

\ifCLASSINFOpdf
   \usepackage[pdftex]{graphicx}
\else
   \usepackage[dvips]{graphicx}
	\usepackage{psfrag} 
\fi

\usepackage[absolute]{textpos}
\newcommand{\copyrightstatement}{
    \begin{textblock}{15}(0.5,0.3)    
         \noindent
         \centering
         \textblockcolour{white}
         \footnotesize
         \copyright 2015 IEEE. Personal use of this material is permitted. Permission from IEEE must be obtained for all other uses, in any current or future media, including reprinting/republishing this material for advertising or promotional purposes, creating new collective works, for resale or redistribution to servers or lists, or reuse of any copyrighted component of this work in other works
    \end{textblock}
}

\begin{document}
%
\title{Estimation of Non-Functional Properties for Embedded Hardware with Application to Image Processing}

\copyrightstatement

\author{
\IEEEauthorblockN{Christian Herglotz, J\"urgen Seiler,  Andr\'e Kaup}
\IEEEauthorblockA{Multimedia Communications and Signal Processing\\
Friedrich-Alexander University Erlangen-N\"urnberg (FAU), \\
Cauerstr. 7, 91058 Erlangen, Germany \\ \{christian.herglotz, juergen.seiler, andre.kaup\}@fau.de}
\and
\IEEEauthorblockN{Arne Hendricks, Marc Reichenbach, Dietmar Fey}
\IEEEauthorblockA{Chair of Computer Architecture\\
Friedrich-Alexander University Erlangen-N\"urnberg (FAU),\\
Martensstr. 3, 91058 Erlangen, Germany \\
{\{arne.hendricks, marc.reichenbach, dietmar.fey\}@cs.fau.de}
}}

\maketitle

\begin{abstract}

In recent years, due to a higher demand for portable devices, which provide
restricted amounts of processing capacity and battery power, the need for energy
and time efficient hard- and software solutions has increased. Preliminary estimations
of time and energy consumption can thus be valuable to improve implementations
and design decisions. To this end, this  paper presents a method to estimate
the time and energy consumption of a given software solution, without having to rely
on the use of a traditional Cycle Accurate Simulator (CAS). Instead, we propose
to utilize a combination of high-level functional simulation with a mechanistic extension
to include non-functional properties: 
Instruction counts from virtual execution are multiplied with corresponding specific energies
and times. 
By evaluating two common image processing algorithms on an FPGA-based CPU, where a mean relative estimation error of $3$\% is achieved for cacheless systems, we show that this estimation tool can be a valuable aid 
in the development of embedded processor architectures. The tool allows the
developer to reach well-suited design decisions regarding the optimal processor hardware
configuration for a given algorithm at an early stage in the design process.

\end{abstract}

\IEEEpeerreviewmaketitle


\section{Introduction}

Recently, the demand for portable devices being capable of performing highly
complex image processing tasks has increased rapidly. Examples are cameras,
smart-phones, and tablet PCs that can capture images and videos in real time.
Furthermore, consumers like to process their data directly to enhance image
quality, compress videos and pictures, or perform other picture manipulating
tasks. Due to their highly complex nature, these tasks are time and energy
consuming which reduces the operating time of the battery significantly. 

Hence, it is desirable to develop energy efficient and fast running image
processing software. For the developer, a major problem in the software design
is that in order to obtain these non-functional properties of an application,
complex and complicated test setups have to be available. E.g., to obtain the
energy consumption of a solution, the code has to be compiled, executed on the
target platform and measured using a power meter. Not till then it is possible
to decide if a solution is energetically efficient. 

To overcome the complex task of measurement we propose to perform a simulation
on a virtual platform that allows to estimate the required energy and time for
the target processing platform. This facilitates predictions on
virtual platforms and allows a very precise estimation.
As in this paper we do not model the cache system, we chose two image processing algorithms as an application.  
These algorithms (video decoding and signal extrapolation) show a highly linear processing order featuring a very high locality such that cache misses play a minor role during execution. The latter algorithm shows a highly homogeneous processing flow using repeatedly the same methods, where the former incorporates highly heterogeneous functions that are called in an unpredictable manner. Hence, we believe that they represent typical algorithms used on portable, battery constrained devices.

By using the open-access platform OVP (Open Virtual Platform) by Imperas it is possible to simulate the execution of such a process on any CPU of interest. 
During the simulation it is counted how often a certain instruction is executed. Multiplying these instruction counts with instruction
specific energies and times we can estimate the complete processing time and
energy of the written code on the desired target platform. In our approach, these specific energies and times are measured beforehand using a predefined set of specialized executable kernels.
By evaluating the influence of an FPU on the chip area, processing time, and energy of the two image processing algorithms we show that our model can help the developer in choosing a suitable processing platform for his application. 
In this contribution, we show that this approach is valid for a cacheless, re-configurable, soft intellectual property (IP) CPU on an FPGA, where further work aims at generalizing this concept to allow the estimation of any CPU.

In this paper, we build upon the work presented in \cite{Berschneider14}. 
We augment this concept by introducing the FPU into the model, showing the viability of this approach for an extended set of test cases, and giving an example for a concrete application. 

The paper is organized as follows: Section \ref{sec:related} presents an
overview of existing approaches  as well as a classification. Section
\ref{sec:OVP} introduces the virtual platform we used to simulate the behavior
of the processor. Subsequently, based on the simulation, the general model is
presented for estimating processing time and energy in Section \ref{sec:model}.
Then, Section \ref{sec:measurements} explains the measurement setup and how the
energies and times for a single instruction as well as for the complete image
processing algorithms is determined. Finally, Section \ref{sec:eval} introduces
two showcase image processing algorithms, evaluates the estimation accuracy for
both these cases and shows how design decisions can be made.

\section{Related Work}
\label{sec:related}

Simulations are crucial tasks when developing hard- and software systems.
Depending on the abstraction level, different simulation methods exist. A
general goal is to simulate as abstract as possible (to enable a fast
simulation) but to be as accurate as needed (to yield the desired properties).
Therefore, we want to discuss several approaches of simulators for micro architectures  and how
to combine them to get very accurate results for non-functional properties like energy and time with a
short simulation time.

The most exact results can be achieved by cycle-accurate simulators
(CAS), such as simulations on hardware description level. These could be RTL
(Register-Transfer-Level) simulations or gate level simulations, where a CAS is
simulating each clock-cycle within a micro architecture. By counting glitches and
transitions, very accurate power estimations can be achieved.
Moreover, multiplying the simulated clock cycles with the clock frequency, the exact execution time can
be determined. Unfortunately, CAS leads to very slow simulation speeds, due to
the fact that they simulate the whole architecture. Typical examples include
Mentor Modelsim, Synopsys VCS and Cadence NCSim.

One possibility to speed up the process is to use SystemC and TLM
(Transaction-Layer-Modeling). By applying the method of \emph{approximately
timed} models, as described in the TLM 2.0 standard, simulations run faster
because clock cycles are combined to so-called \emph{phases}. However, in
contrast to the CAS simulations, the counted times and transitions are less
exact. A simulator which follows this paradigm is the SoCLib library
\cite{Elabidine14}. Moreover, frameworks such as Power-Sim have been developed
to speed-up the simulation time by providing quasi cycle-accurate simulation.

On a higher abstraction level, simulation tools like Gem5 combined with external
models like Orion or McPAT \cite{Binkert} can measure both time and energy 
to a certain extent, while at the same time sacrificing simulation performance:
Complex applications like image processing can take up to several days until
simulation is finished. Thus, if only functional properties such as correctness
of algorithm, results or completion are of interest, an instruction accurate
simulator is needed in order to speed up the simulation process. 

Instruction set simulators (ISS) do not simulate the exact architecture, but
rather ``interpret'' the binary executable and manage internal registers of the
architecture. Typically, they require the least simulation time, but on the
other hand do not include  non-functional results such as processing time and energy
consumption \cite{Sanchez}. They are usually used to emulate systems which
are not physically present, as debugger or as development platform for embedded
software to get functional properties. OVPSim can be mentioned as an example for
this simulator class.

The discussed approaches show that a very abstract description-level usually
goes hand-in-hand with less information to be retrieved from simulation. 
One way to overcome this problem is executing on a very abstract level but
compensating inaccurate results by applying a mathematical or statistical model in
order to obtain relatively accurate estimations.

A typical example for this approach is presented by Carlson et al.: The Sniper
simulator. According to \cite{Carlson}, it is a simulator for x86 architectures
such as Intel Nehalem and Intel Core2. For an interval model
\cite{Eeckhout}, where an interval is a series of instructions, possible cache and branch misses (due to interaction) are estimated. With the help of this information, homogeneous and heterogeneous desktop multi-core
architectures are simulated. This is helpful when modeling rather complicated desktop
architectures that include
out-of-order-execution, latency hiding and varying levels of parallelism
\cite{Eyermann}. While they are able to simulate multi program workloads and
multi threaded applications, one drawback is the relative error which rises up to $25\%$, which is
acceptable for energy uncritical desktop applications and multi program
simulation, but unacceptable when trying to find optimal design choices for energy aware embedded hardware.  Furthermore, SniperSim is solely
focused on Intel desktop architectures, which makes its use for general embedded
hardware simulation unfeasible.

\begin{figure}[tbp]
\begin{center}
	\includegraphics[scale = .35]{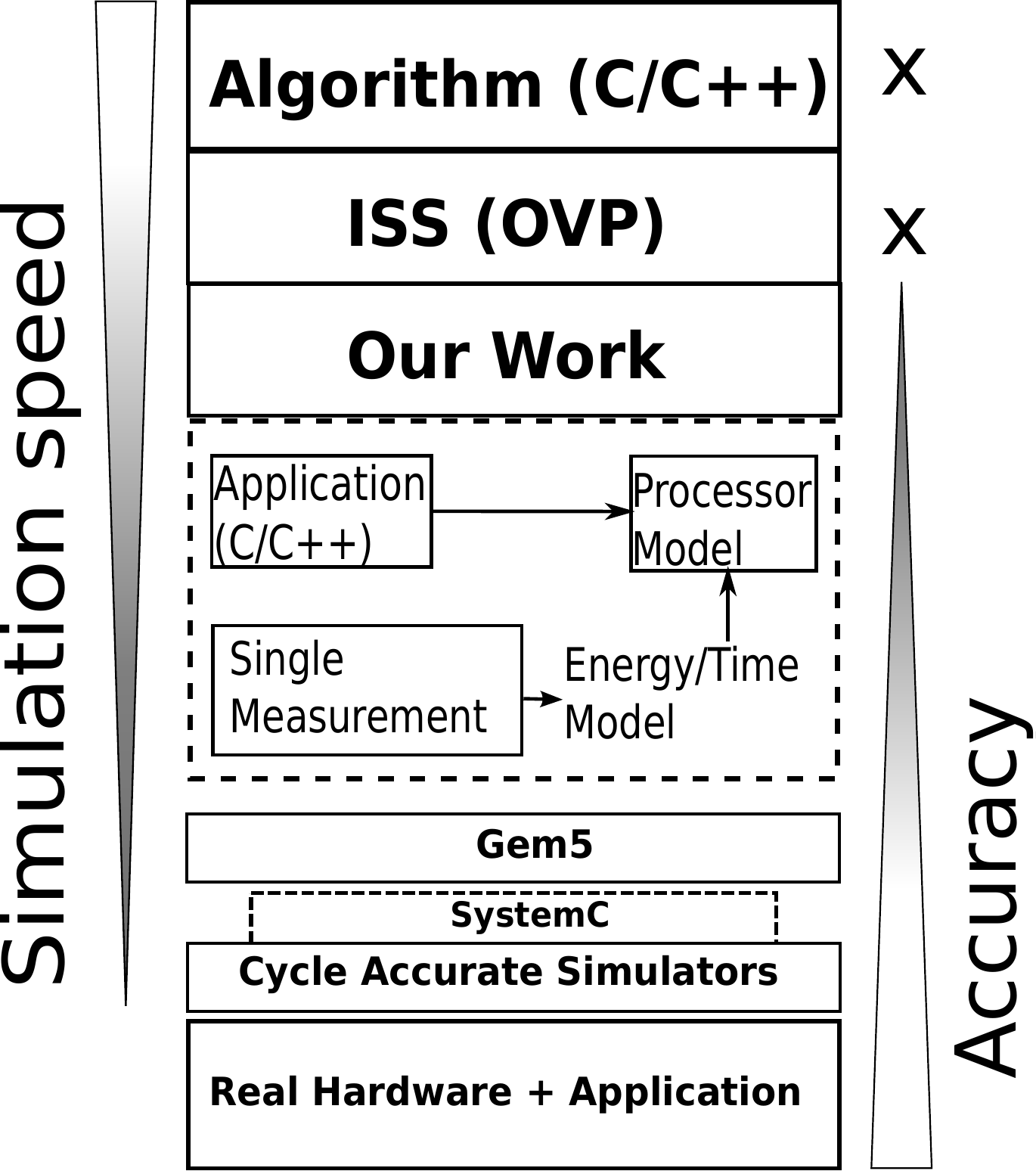}
	\caption{Different simulation tools regarding simulation speed (left) and estimation accuracy (right)
   in regard to non-functional properties. All the tools provide functional
   parameters, non-functional properties cannot be obtained by ISS or the algorithm.} 
    	\label{simulation_tools}
\end{center}
\end{figure}

An illustration of the presented different simulation layers (including our extension of an
ISS), ranged by their respective simulation speed, as well as the accuracy of the
resulting non-functional parameter estimations, can be seen in Figure
\ref{simulation_tools}. 
Summarizing, we propose to estimate the processing time and energy using an extended ISS with only slightly increased simulation times. 
Therefore, we chose an existing 
tool that features a wide variety of embedded architectures, the OVP framework.  Efforts
have been undertaken to find timing models for CPUs  modeled in OVP in the past
\cite{Rosa}, where the authors extended parts of a framework with pseudo-cycle
accurate timing models. Using a watchdog component, they
integrated an assembly parser and hash table with pre-characterized groups of
instruction and timing information. The component then analyzes every
instruction based upon a disassembly by OVP and assigns a suitable
timing information.  Unfortunately, simulation run-time is poor due to the external analysis of the instructions via disassemblation. 
In regards of power modeling, Shafique et al. \cite{Shafique14} proposed an adaptive management system for
dynamically re-configurable processors that considers leakage and dynamic energy.
Our approach is different as our emphasis is not on a low-level estimation
including register-transfer level (RTL) techniques such as power-gating or instruction set building (often
leading to a long simulation run-time),  but rather on a fast high-level mechanistic simulation. 
Further work in regards of power and
time consumption has also been done by Gruttner et al. in \cite{Nebel14}. Their
focus, however, lies on rapid virtual system prototyping of SoCs using C/C++
generated virtual executable prototypes utilizing code annotation.

In order to motivate our approach we first explain the fundamentals of a
mechanistic simulation \cite{Eeckhout}, which can be employed when using an ISS.
Mechanistic simulation means simulating bottom up, starting from a basic
understanding of an architecture. In this context, the source of
this understanding is unimportant, thus it can also be achieved by an empirical
training phase, which makes it suitable to proprietary IPs with limited
knowledge of underlying details. 
This can be done by running measurements or experiments on actual hardware, in
our case measuring processing energy and time of instructions. The resulting data is
then prepared to be used in the simulation model - preparation  can
include regression and fitting of parameters.  A mechanistic simulation is then
run on a typical set of instructions using constant costs per instruction. In a very early work, this
concept has been analyzed by Tiwari et al. \cite{Tiwari94} by measuring the
current a processor consumes when a certain instruction is executed. Our approach 
 is presented in the next section. 

 
\section{Instruction Set Simulator: Virtual Platforms} 
\label{sec:OVP}

As described in the section before, in this paper we want to combine an
ISS with an additional model to get non-functional
properties such as energy consumption and processing time. To achieve this, we
discuss the ISS in detail and explain how
it is modified for our purposes.

Open Virtual Platform (OVP) is a functional simulation environment which
provides very fast simulations even for complex applications. Moreover, this
flexible simulation environment is easy to use because once compiled
applications can be run on both the real hardware and the OVP simulator without
additional annotations to the program. The fast simulation run-times even for
complex application is possible because OVP simulates instruction-accurate, not
cycle-accurate processing. The user can debug the simulation and will know at any point,
e.g., which content the registers and program counters have, but not the current
state of the processor pipeline. Analyzing non-functional properties like energy
consumption or execution run-time are therefore natively not
possible \cite{OVPWhitePaper}. 

To run a simulation using OVP two things are necessary, first a so-called
platform model which denotes the current hardware platform, e.g., a CPU and a
memory. Second, the application has to be  provided as a binary executable file
(the kernel).  For a fast start in OVP, several processor and peripheral models
are included, where some of them are open source (e.g., the OR1K processor
model). These models are dynamically linked during run-time to the simulator
called OVPsim. Today, there is a number of different manufacturers of embedded
processors on the market, such as ARM or INTEL. Because they are Hard-IP (intellectual property), these
processors have the disadvantage that they cannot be individualized to the needs
of the hardware designer, e.g., grade of parallelism, cache size or cache
replacement strategy. Fortunately, there are also some open source Soft-IP processors available, like the
LEON3 processor \cite{LEON3}. 
This processor can
be edited individually to the needs of the hardware designer. The LEON3
implements the SPARC V8 processor architecture, which is available as VHDL
source code.  The different configurations of the processor can be synthesized
and tested on an FPGA. 

For our work, we have chosen the LEON3 processor because of its configurability.
It is easier to analyze energy and time if unnecessary components can be
disabled to allow a well defined measurement environment.  Moreover, the
availability as an open source soft IP core allows easy debugging. As previously
there was no SPARC V8 processor model available, a new complete processor model
was developed by us~\cite{berschneiderBSC}. By that we reach the same level of
flexibility and configuration possibilities on simulation as in the processor on
real hardware. Therefore, a C API for implementing and extending own processor
models provided by OVP was used.

\begin{figure}[tbp]
\begin{center}
	\includegraphics[width=0.45\textwidth]{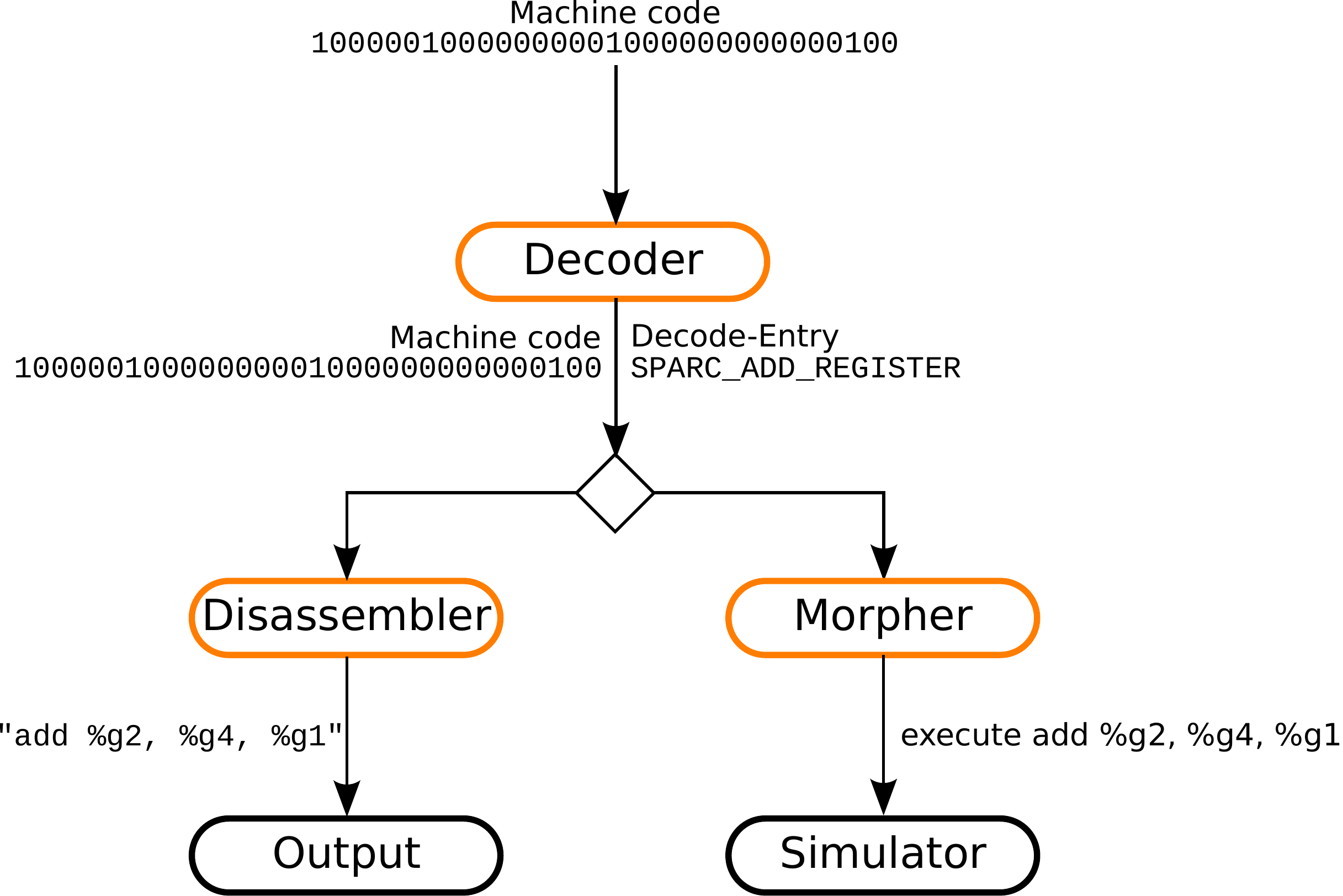}
	\caption{Simulation-Workflow of an instruction in OVP. }
    	\label{cpumodel_flow}
\end{center}
\end{figure}

The general simulation flow of a single instruction in our SPARC V8 processor
model is visualized in Fig. \ref{cpumodel_flow}. First, the decoder
analyzes the 32-bit instruction for patterns and decides what kind of instruction it
is. Then, the instruction receives an internal tag which is used for representation in the
disassembler and morpher. The disassembler includes functions for simulation
output if the user wishes to debug the simulated instructions. The morpher part
of the processor model generates native code for the simulator to execute. These
functions represent what the simulator should do. 
E.g., an arithmetic operation
extracts the source registers, reads the values of these registers, executes the
operation, and saves the result to the target register. Moreover, depending on the
kind of arithmetic operation, the internal ALU state can be changed to implement further instructions like branches. 

\begin{figure}[tbp]
        \begin{center}
    \includegraphics[width=0.36\textwidth]{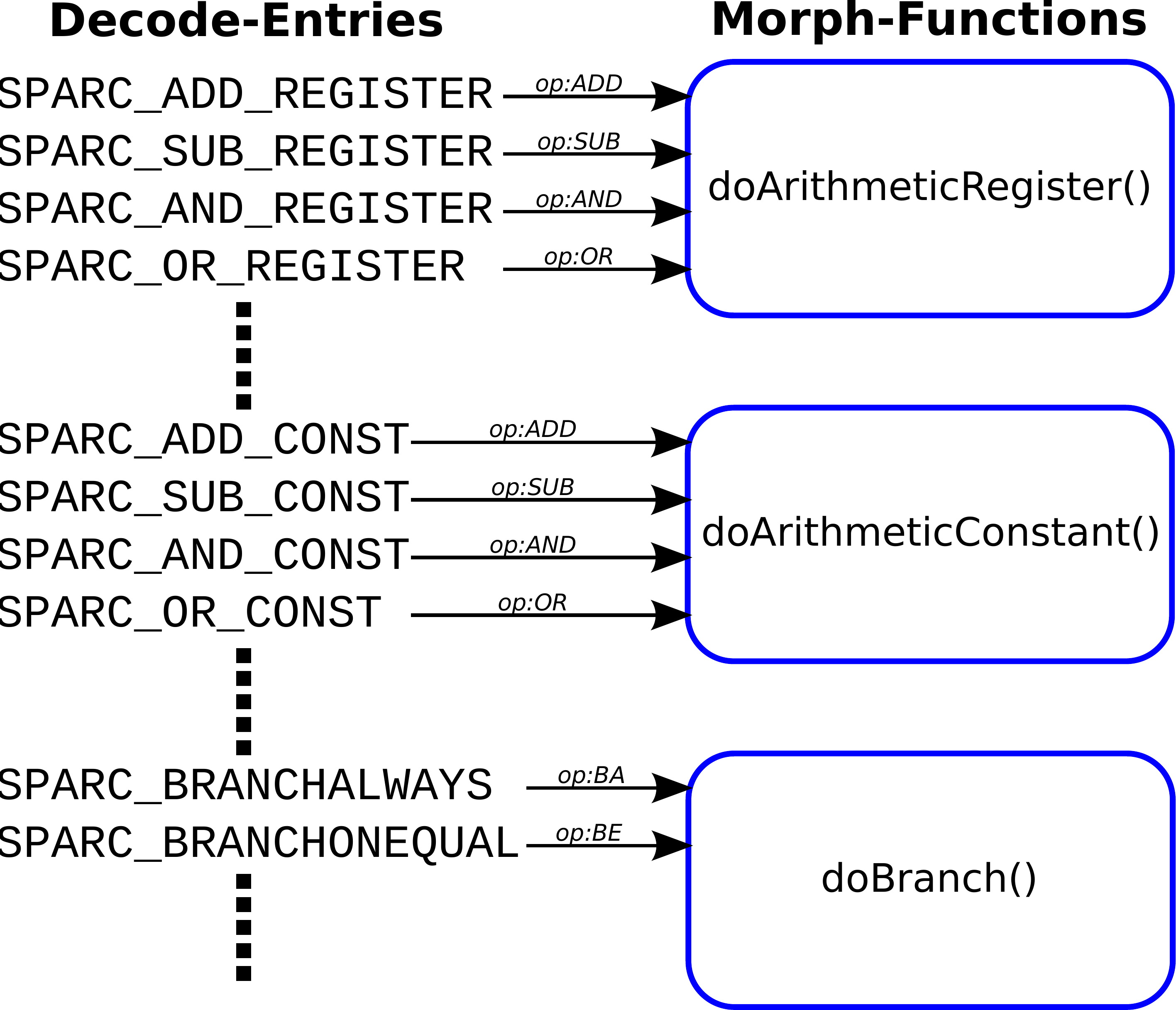}
	\caption{Allocation from Decode-Entries to Morph-Functions, which create native code for the simulator. }
    	\label{decodeentries-morphfunctions}
\end{center}
\end{figure}

As writing a morpher function for every possible instruction is highly complex, instructions were grouped and morphing functions
summarized. E.g., arithmetic instructions like \texttt{add} or \texttt{sub} and their
variants (analyzing flags, setting flags) are one group. Because of
different data manipulation, register-register and register-immediate
instructions had their own group, e.g., arithmetic-register-register
instructions and arithmetic-register-immediate instructions. Figure
\ref{decodeentries-morphfunctions} shows a visualization of this grouping.

The methods described above were used to get a fully functional simulation
environment. To enable the estimation of non-functional properties, the
functional simulation has to be extended. On real hardware, not all instructions
have the same data or control path in the processor, e.g., a floating point
operation is much more complex, needs more cycles and therefore more run-time
than a simple integer instruction.  Thus, all instruction groups are further divided into
categories like integer, floating point, jumps, etc. The internal counters for these instruction categories are realized
without using callback functions to ensure a high simulation speed. Instead, in
every morpher function a counter is implemented that increases an internal
temporal register after the corresponding instruction was executed by the
simulator. For every category one internal register exists. After the
full execution of the application, the simulator reads out these registers and
presents the results.

In this context we want to mention that the non-functional behavior of an instruction is not necessarily constant. Especially for complex instruction set architectures (CISC) the context has a major impact on energy and time 
(instructions can take longer due to mispredicted branches,
depending on where they are found in the context of the program). On the other hand, for reduced instruction set architectures (RISC), most instructions can be executed using fewer cycles compared to a CISC based system.
Time and energy waste for, e.g., a flushed pipeline due to a mispredicted branch are
consequently not as severe when compared to a CISC based processor. 
As our work is mainly focused on embedded hardware,
which often implements RISC architectures such as the Sparc
V8 based LEON3 (which does not even feature a pipeline), we argue that due to 
architectural properties it is valid to assume that an instruction
shows a roughly constant non-functional behavior regardless of the context it is found in.

\section{Energy and Time Modeling}
\label{sec:model}
The general equations used to estimate the processing energy $\hat E$ and time $\hat T$ are given as 
\begin{equation}
        \hat E = \sum_c e_c \cdot n_c \quad \text{and} \quad
   \hat T = \sum_c t_c \cdot n_c,
   \label{eq:energy}
\end{equation}
where the index $c$ represents the instruction category as introduced before, $e$ and $t$ the instruction specific energy and time, and $n$ the instruction count. 

For our model, nine instruction categories have been identified as summarized in Table \ref{tab:instCategs}. The first six categories describe the energy consumption of the basic integer unit. The remaining three categories correspond to the FPU operations. 
The category ``FPU Arithmetic'' comprises the floating point add, subtract, and multiply operation. 

\begin{table}
\centering
\renewcommand{\arraystretch}{1.3}
\caption{Instruction categories and their respective specific energies and times as derived by dedicated measurements.  }
\label{tab:instCategs}
\begin{tabular}{l|r|r}
\hline
Instruction category $c$ & Spec. Time $t_c$ & Spec. Energy $e_c$ \\
\hline
Integer Arithmetic 	& $45$ns 	& $15$nJ\\
Jump 				& $238$ns 	& $76$nJ\\
Memory Load 		& $700$ns	& $229$nJ\\
Memory Store 		& $376$ns 	& $166$nJ\\
NOP 				& $46$ns 	& $13$nJ \\
Other 				& $41$ns 	& $13$nJ \\
FPU Arithmetic 		& $46$ns 	& $14$nJ\\
FPU Divide 			& $431$ns 	& $431$nJ\\
FPU Square root	 	& $612$ns 	& $88$nJ\\
\hline
\end{tabular}
\end{table}

The middle and the right column of Table \ref{tab:instCategs} include the
instruction specific energies and times that are assigned to the respective
categories. The specific time $t_c$ can be interpreted as the mean time required
to execute one instruction of this category in our test setup. Likewise, the
specific energy $e_c$ describes the mean energy needed during the execution of
one instruction. The values shown in the table have been derived by the
measurement method explained in Section \ref{sec:measurements}. 

Now if we know how often an instruction of a given category is executed during a
process, we can multiply this number with the specific energy and time, add up
the accumulated values for each category and obtain an estimation for the
complete execution time and energy. 
These numbers $n_c$ are called \emph{instruction counts}
and they are derived by the simulation in the ISS as presented above. 

\section{Measurement Setup}
\label{sec:measurements}

In order to prove the viability of the presented model, we built a dedicated test
setup for measuring the execution time and energy consumption of a SPARC LEON3
softcore processor \cite{LEON3} on an FPGA board. The FPGA board was a Terasic
DE2-115 featuring an Altera Cylcone IV FPGA. The board was controlled using
GRMON debugging tools \cite{grmon2} and the LEON3 was synthesized using Quartus,
where the cache system and the MMU were disabled. Hence, in this publication, we
consider a baseline CPU including an FPU. 
We utilized an FPGA because it offers great flexibility: The CPU can be customized according to our needs for highly versatile testing. We exploited this property to generate a useful platform for the step-by-step construction of an accurate and general RISC model.


For the measurement of the execution time of a process we used the \texttt{clock()}-function
from the C++  standard library \texttt{time.h}. The measurement method for the energy consumption of the process is the same as already presented in \cite{Berschneider14}. 

To obtain the energy required to execute a single instruction we measured two kernels: A reference and a test kernel as indicated
by Table \ref{tab:fileSetup}. The processing in both kernels features the same
amount of baseline instructions, e.g., jumps for a loop. In contrast, the test kernel additionally contains a high amount of
specific instructions that are not included in the reference kernel. Subtracting
the processing energy and time of the reference kernel ($E_\mathrm{ref}$, $T_\mathrm{ref}$) from that of the test
kernel ($E_\mathrm{test}, T_\mathrm{test}$), we obtain the time and energy required by the additional instructions. This value is then divided by the number of instruction executions $n_\mathrm{test}$, which is the product of the number of loop cycles with the number of instructions inside the loop. Thus, we obtain an instruction specific time and energy as 
\begin{equation}
e_c = \frac{E_\mathrm{test}-E_\mathrm{ref}}{n_\mathrm{test}} \quad \mathrm{and} \quad t_c = \frac{T_\mathrm{test}-T_\mathrm{ref}}{n_\mathrm{test}}. 
\end{equation}
Due to the unrealistic programming flow of the reference and the test kernel, these values may differ from the values observed in a real application. Hence, the values are checked for consistency and manually adapted, if necessary. 

It should be noted that the energy for the execution of a certain instruction can be variable depending on the preceding and succeeding instruction, or the input data. To overcome this problem, we take the assumption that in real application, this variation is averaged to an approximately constant value when the corresponding instruction is executed multiple times in different contexts, which is supported by the results of our evaluation.

\begin{table}
\centering
\renewcommand{\arraystretch}{1.3}
\caption{Pseudo-code of reference and test file to obtain time- and energy specific values. The reference file contains a \texttt{for} loop without any content. In the test file, the \texttt{for} loop contains a large amount of the instructions to be tested, in this example integer add operations.  }
\label{tab:fileSetup}
\begin{tabular}{p{0.24\textwidth} |p{0.24\textwidth}}
\texttt{\slash\slash ~ Reference kernel }& \texttt{\slash\slash ~ Test kernel} \\
\texttt{int main(.)} & \texttt{int main(.)}\\
\texttt{\{} & \texttt{\{}\\
\texttt{\quad for (i=0; i<1000000; i++) } & \texttt{\quad for (i=0; i<1000000; i++) }\\
\texttt{\quad \{} & \texttt{\quad\{}\\
\texttt{\qquad \slash\slash ~ empty } & \texttt{\qquad ADD \# \#}\\
\texttt{ } & \texttt{\qquad ...}\\
\texttt{ } & \texttt{\qquad ADD \# \#}\\
\texttt{\quad \}} & \texttt{\quad\}}\\
\texttt{\}} & \texttt{\}}\\

\end{tabular}
\end{table}


\section{Evaluation}
\label{sec:eval}
In this section, we show that our model returns valid energy and time estimations for the given CPU by testing two conventional image processing algorithms: High-Efficiency Video Coding (HEVC) decoding that performs mainly integer arithmetics and Frequency Selective Extrapolation (FSE) which makes extensive use of floating point operations. 

\subsection{HEVC Decoding}
To test the energy consumption of the HEVC decoder we used the HM-reference software \cite{HM-11.0} that was slightly modified to run bare-metal. Therefore, we included in- and output streams directly into the kernel and cross-compiled it for the LEON3. While the HEVC can be fully implemented using pure integer arithmetics, the used software performs few floating point operations, e.g., for timing purposes. Furthermore, it uses a high variety of different algorithmic tools and methods like filtering operations and transformations. Due to predictive tools, a high amount of memory space is required. 

To have a representative test set, we measured the decoding process of $36$ different video bit streams. These bit streams were encoded with four different encoding configurations (intra, lowdelay, lowdelay\_P, and randomaccess), three different visual qualities (quantization parameters 10, 32, and 45), and three different input raw sequences. 

\subsection{Frequency Selective Extrapolation}
The second test of the model was carried out by computing the Frequency Selective Extrapolation (FSE) \cite{Seiler2010} algorithm on the device. FSE is an algorithm for reconstructing image signals which are not completely available, but rather contain regions where the original content is unknown. This may, e.g., happen in the case of transmission errors that have to be concealed or if an image contains distortions or undesired objects. For the extrapolation purpose, FSE iteratively generates a parametric model of the desired signal as a weighted superposition of complex-valued Fourier basis functions. As the model is defined for the available as well as for the unknown samples, one directly obtains an extension of the signal into these unknown regions.

For generating the model, the available samples are iteratively approximated at which in every iteration one basis function is selected by performing a weighted projection of the approximation residual on all basis functions. This process can also be carried out in the frequency-domain. In doing so, a Fast Fourier Transform is necessary. Due to the high amplitude range and the required accuracy, all operations need to be carried out with \texttt{double} precision accuracy. For a detailed discussion of FSE, especially the relationship between the spatial-domain implementation and the frequency-domain implementation and its influence on the required operations and run-time, please refer to \cite{Seiler2010, Seiler2011}.

As a test set for the input data we chose 24 different pictures from the Kodak test image database where for each picture a different mask was defined. Hence, we obtained $24$ kernels differing by the input image and mask. 

\subsection{Experimental Results}
To see the benefit and the influence of an additional FPU in a CPU, for both algorithms we tested two cases: Processing with and without floating point operations (float and fixed, respectively). The latter case is achieved by compiling the kernel using the compiler flag \texttt{-msoft-float} which emulates floating point operations using integer arithmetics. The use of this compiler flag does not influence the precision of the process such that the output matches exactly the output of the kernel that was compiled without this flag. 

\begin{figure}
\centering
\ifCLASSINFOpdf
\else
\psfrag{000}[c][c]{\footnotesize{$0$}}
\psfrag{001}[c][c]{\footnotesize{$5$}}
\psfrag{002}[c][c]{\footnotesize{$10$}}
\psfrag{003}[c][c]{\footnotesize{$15$}}
\psfrag{004}[c][c]{\footnotesize{$20$}}
\psfrag{005}[c][c]{\footnotesize{$14.3$}}
\psfrag{006}[c][c]{\footnotesize{$14.35$}}
\psfrag{007}[c][c]{\footnotesize{$14.4$}}
\psfrag{008}[c][c]{\footnotesize{$14.45$}}
\psfrag{009}[l][l]{\footnotesize{Test file}}
\psfrag{010}[l][l]{\footnotesize{Reference file}}
\psfrag{011}[c][c]{\footnotesize{Measured Power [$W$]}}
\psfrag{012}[c][c]{\footnotesize{Time [$s$]}}
\fi
\ifCLASSINFOpdf
\vspace{-4.6cm}
\includegraphics[width=3.6in]{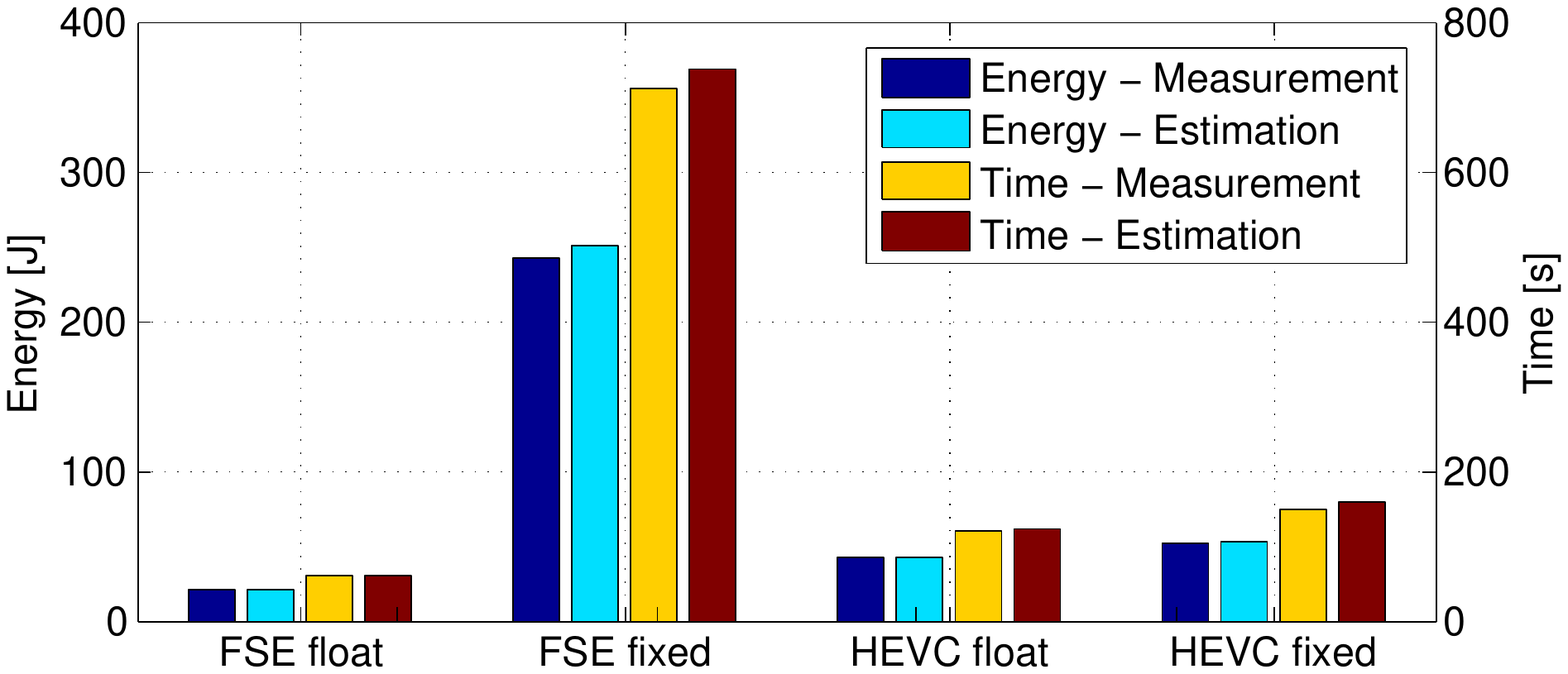}
\vspace{-5cm}
\else
\includegraphics[width=3in]{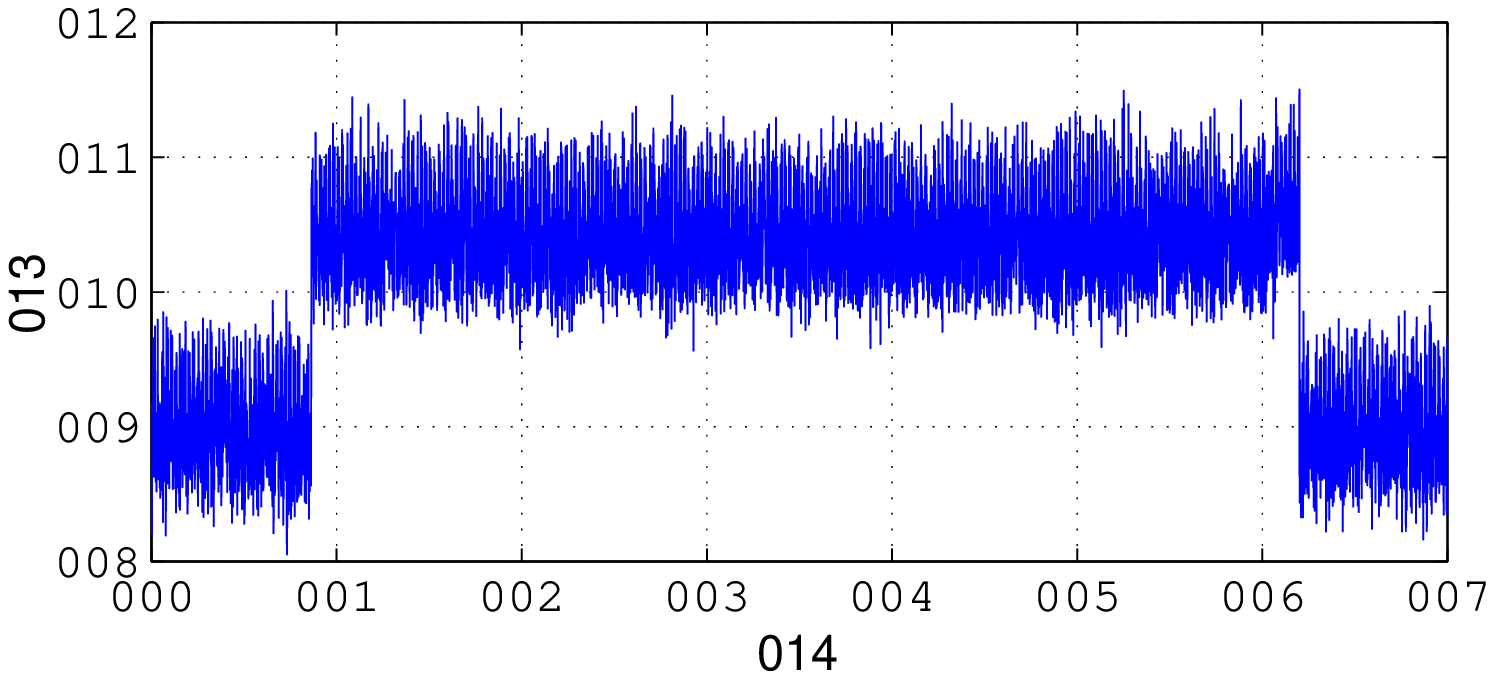}
\fi
\caption{Comparison between measurement and estimation for four different showcase processes. The two FSE kernels as well as the two HEVC kernels process the same input data. In contrast to the float kernels, the fixed kernels are compiled with the \texttt{-msoft-float} compiler flag. }
\label{fig:comparison_est_meas}
\end{figure}

We show the validity of the model by measuring the execution time and the energy consumption of the processes presented above and comparing them to the estimations returned by the proposed model. The bar diagram in Figure \ref{fig:comparison_est_meas} shows the results for four different representative cases. 

The dark blue bars depict the measured energies, the light blue bars the estimated energies (left axis). The yellow bars represent the measured times and the red bars the estimated times (right axis). We can see that all estimations are located close to their corresponding measured values. 

To evaluate the performance of our algorithm, we calculated the estimation errors for all $M$ tested kernels as 
\begin{equation}
\varepsilon_m = \frac{\hat E_m - E_{\mathrm{meas},m}}{E_{\mathrm{meas},m}}, 
\end{equation}
where $\hat E_m$ is the estimated energy from (\ref{eq:energy}), $E_{\mathrm{meas},m}$ is the measured energy, and $m$ is the kernel index. In the same way, the estimation error for time was derived. Table \ref{tab:errors} shows two summarizing indicators for the evaluated kernels: first, the mean absolute estimation error $\bar \varepsilon = \frac{1}{M}\sum_{m=1}^M \left| \varepsilon_m\right|$ 
and second, the maximum absolute error $\varepsilon_\mathrm{max} = \max_m \left| \varepsilon_m \right|,\quad m=1,...,M$ for both energy and time. The maximum error is the highest error we observed for our evaluated kernel set. 
\begin{table}
\centering
\renewcommand{\arraystretch}{1.3}
\caption{Mean absolute estimation error and maximum absolute error of our model.  }
\label{tab:errors}
\begin{tabular}{l|r|r}
\hline
 & Energy & Time \\
\hline
Mean absolute error $\bar \varepsilon$ 	& $2.68\%$ 	& $2.72\%$\\
Maximum absolute error $\varepsilon_\mathrm{max}$	& $6.32\%$ 	& $6.95\%$\\
\hline
\end{tabular}
\end{table}
The small mean errors which are lower than $3\%$ show that the estimations of our model can be used to approximate the real energy consumption and processing time. 

\subsection{Application of the Model}
As a first basic application of this model the above presented information can be used to help the developer decide for a suitable architecture. If he, e.g., would like to know if it makes sense to include an FPU on his hardware, he can simulate the execution of his code with and without an FPU and obtain the information about processing time and consumed energy. The result of such a benchmark for our framework is shown in Table \ref{tab:gains}. 

\begin{table}
\centering
\renewcommand{\arraystretch}{1.3}
\caption{The change of non-functional properties of an algorithm when introducing an FPU into the hardware. }
\label{tab:gains}
\begin{tabular}{l|r|r}
\hline
 & FSE & HEVC Decoding \\
\hline
Energy consumption 	& $-92.6\%$ 	& $-42.88\%$\\
Processing Time 	& $-92.8\%$ 	& $-43.49\%$\\
\# logical elements 	& $+109\%$ 	& $+109\%$\\
\hline
\end{tabular}
\end{table}

The values in the table are mean values over all tested kernels. The third row shows the increase of chip area needed for an FPU which can be obtained by the synthetization of the processor. We can see that if we spend more chip area (about twice the size as indicated by the number of logical elements), we save more than $90\%$ of the processing time and energy for FSE processing, which may be highly beneficial for energy and time constrained devices. In contrast, for HEVC decoding, an FPU reduces time and energy by not even half such that the expensive chip area might lead the developer to choose for a processor without an FPU. 

\section{Conclusions}
In this paper, we presented a method to accurately estimate non-functional properties of an algorithm using simulations on a virtual platform. The model is based on a mechanistic approach and reaches an average error of $2.68\%$ for energy and $2.72\%$ for time estimation. Furthermore, we have shown that the information can be used by developers for time and energy estimates. Further work aims at incorporating a model for the cache and multi-core processors and generalizing this concept to any CPU of interest. Additionally, we will evaluate the estimation accuracy of this model for further algorithms to show the general viability.

\section*{Acknowledgment}
This work was financed by the Research Training Group 1773 ``Heterogeneous Image Systems'', funded by the German Research Foundation (DFG).

%
\bibliographystyle{abbrv}
\bibliography{literature}  

\begin{thebibliography}{10}

\bibitem{LEON3}
Leon3 processor.
\newblock [Online.] Available:
  \url{http://gaisler.com/index.php/products/processors/leon3?task=view&id=13}.

\bibitem{OVPWhitePaper}
B.~Bailey.
\newblock System level virtual prototyping becomes a reality with {OVP}
  donation from imperas.
\newblock {\em White Paper, June}, 1, 2008.

\bibitem{berschneiderBSC}
S.~Berschneider.
\newblock Modellbasierte hardwareentwicklung am beispiel eingebetteter
  prozessoren f\"ur die optische messtechnik.
\newblock Master's thesis, Friedrich-Alexander-Universit\"at
  Erlangen-N\"urnberg, Germany, Dec. 2012.

\bibitem{Berschneider14}
S.~Berschneider, C.~Herglotz, M.~Reichenbach, D.~Fey, and A.~Kaup.
\newblock Estimating video decoding energies and processing times utilizing
  virtual hardware.
\newblock In {\em Proc. 3PMCES Workshop. Design, Automation \& Test in Europe
  (DATE)}, 2014.

\bibitem{Binkert}
Binkert and Beckmann.
\newblock The gem5 simulator.
\newblock {\em ACM SIGARCH Computer Architecture News}, 39(2):1--7, May 2011.

\bibitem{Carlson}
T.~E. Carlson, W.~Heirman, and L.~Eeckhout.
\newblock Sniper: Exploring the level of abstraction for scalable and accurate
  parallel multi-core simulations.
\newblock In {\em Proc. International Conference for High Performance
  Computing, Networking, Storage and Analysis}. ACM, Nov 2011.

\bibitem{Eeckhout}
L.~Eeckhout.
\newblock Computer architecture performance evaluation methods.
\newblock {\em Synthesis Lectures on Computer Architecture}, 5:1--145, 2010.

\bibitem{Elabidine14}
K.~Z. Elabidine and A.~Greiner.
\newblock An accurate power estimation method for {MPSoC} based on {SystemC}
  virtual prototyping.
\newblock In {\em Proc. 3PMCES Workshop. Design, Automation \& Test in Europe
  (DATE)}, 2014.

\bibitem{Eyermann}
S.~Eyerman, L.~Eeckhout, T.~Karkhanis, and J.~E. Smith.
\newblock A mechanistic performance model for superscalar our-of-order
  processors.
\newblock {\em ACM Transactions on Computer Systems (TOCS)}, 27(2), May 2009.

\bibitem{grmon2}
A.~Gaisler.
\newblock Grmon2 debug monitor.
\newblock [Online.] Available:
  \url{http://www.gaisler.com/index.php/products/debug-tools/grmon2}.

\bibitem{Nebel14}
K.~Gruttner, P.~Hartmann, T.~Fandrey, K.~Hylla, D.~Lorenz, S.~Stattelmann,
  B.~Sander, O.~Bringmann, W.~Nebel, and W.~Rosenstiel.
\newblock An {ESL} timing and power estimation and simulation framework for
  heterogeneous {SoC}s.
\newblock {\em Proc. International Conference on Embedded Computer Systems:
  Architectures, Modeling, and Simulation (SAMOS XIV)}, pages 181--190, 2014.

\bibitem{HM-11.0}
{ITU\slash ISO\slash IEC}.
\newblock {HEVC Test Model HM-11.0}.
\newblock [Online.] Available: \url{https://hevc.hhi.fraunhofer.de/}.

\bibitem{Rosa}
F.~Rosa, L.~Ost, R.~Reis, and G.~Sasatelli.
\newblock Instruction-driven timing {CPU} model for efficient embedded software
  development using {OVP}.
\newblock In {\em Proc. IEEE International Conference on Electronics, Circuits,
  and Systems (ICECS)}, 2011.

\bibitem{Sanchez}
D.~Sanchez and C.~Kozyrakis.
\newblock Zsim: Fast and accurate microarchitectural simulation of
  thousand-core systems.
\newblock {\em Proc. 40th Annual International Symposium on Computer
  Architecture (ISCA)}, 2013.

\bibitem{Seiler2010}
J.~Seiler and A.~Kaup.
\newblock Complex-valued frequency selective extrapolation for fast image and
  video signal extrapolation.
\newblock {\em IEEE Signal Processing Letters}, 17(11):949 -- 952, November
  2010.

\bibitem{Seiler2011}
J.~Seiler and A.~Kaup.
\newblock A fast algorithm for selective signal extrapolation with arbitrary
  basis functions.
\newblock {\em EURASIP Journal on Advances in Signal Processing}, 2011:1--10,
  2011.

\bibitem{Shafique14}
H.~Shafique, Bauer.
\newblock Adaptive energy management for dynamically reconfigurable processors.
\newblock {\em IEEE Transactions on Computer-Aided Design of Integrated
  Circuits and Systems}, 33(1):50--63, January 2014.

\bibitem{Tiwari94}
V.~Tiwari, S.~Malik, and A.~Wolfe.
\newblock Power analysis of embedded software: A first step towards software
  power minimization.
\newblock {\em IEEE Transactions on VLSI Systems}, 2(4):437--445, Dec 1994.

\end{thebibliography}

\end{document}